\documentclass{article}
\usepackage{cite}
\usepackage{amsmath,amssymb,amstext,times}
\usepackage{color}
\usepackage{float}
\usepackage{pdfpages}
\usepackage{tcolorbox}
\usepackage{tabularx}
\usepackage{array}
\usepackage{colortbl}
\usepackage{arydshln}
\usepackage{array,multirow}
\usepackage{setspace} % for figures text
\tcbuselibrary{skins}

\title{Experimental similarity assessment for a collection of fragmented artifacts}

 \author{S. Biasotti, E. Moscoso Thompson and M. Spagnuolo}
\begin{document}
	
\maketitle
\begin{abstract}
In the Visual Heritage domain, search engines are expected to support archaeologists and curators to address cross-correlation and searching across multiple collections. %To this purpose, search engine must be flexible, multi-modal and support qualitative and quantitative queries. 
Archaeological excavations return artifacts that often are damaged with parts that are fragmented in more pieces or totally missing. The notion of similarity among fragments cannot simply base on the geometric shape but style, material, color, decorations, etc. are all important factors that concur to this concept.
In this work, we discuss to which extent the existing techniques for 3D similarity matching are able to approach fragment similarity, what is missing and what is necessary to be further developed.
\end{abstract}
%-------------------------------------------------------------------------

\section{Introduction}
In the Cultural Heritage domain, the digital era makes available scans of artifacts showing their 3D geometry, photometric data, chemical properties and also digitized card catalogue information on provenance, classification, and full-text archaeological descriptions from various sources. The availability of this wealth of data and information opens-up the possibility to build new tools to study and explore artifacts, leveraging on the digital accessibility to large sets of exemplars  that can provide insights difficult to achieve by manual inspection of artifacts.

Visual search engines are the natural tool to consider for supporting comparative studies of CH artifacts, but the typical shape similarity measures used in visual search engines are not sufficient to characterize the many different flavors of shape comparison useful in the CH domain. Search should be flexible, multi-modal, should support cross-correlation and searching across collections. This is particularly important when the assets to be studied are only parts of objects, and therefore their overall shape does not have in most cases a meaning.

%\textcolor{red}{Alcuni punti per un'introduzione... qui vorrei dire, come ci piazziamo nel contesto CH, cosa facciamo in questo lavoro in termini di descrittori e classificazione del dataset, come valutiamo... anche se non riesco a immaginarmi molto di più che un'analisi visuale...}

% Cultural Heritage users are not experts in Computer Graphics nor in Computer Vision nor in Information Retrieval and, in general, users to not have one specific similarity search in mind, rather, they need various criteria to look for various similarities among fragments.

In this paper, we present the settings and initial results of an experimentation carried out in the EU GRAVITATE project \cite{GRAVITATE_prog} on datasets of fragments of terracotta statues, jars and other archaeological finds which have been selected as case studies in the project. Beside the geometry, fragments are digitized with colorimetric information on the mesh vertices, and fully documented by curatorial data. The integration of visual search with metadata and semantic search is not discussed in this paper and further details can be found \cite{GRAVITATE_prog,dashboard}.

The main goal of this experimentation is to understand at what extent existing similarity techniques are able to support archaeologists in the their cataloging and classification work
%the intuitive concept of similarity that underlies CH's studies can be replicated by existing techniques, 
and how these techniques are able to answer to the fragment similarity problem, discussing what is currently missing and what is necessary to be further developed.

One critical point when dealing with similarity assessment among archaeological fragments is that there is not a unique possible similarity criterion, but different aspects become relevant depending on what is the target of the similarity reasoning, and therefore, what should be emphasized. For instance, the fragment's thickness becomes an important search cue if the target is the reassembly with other parts, or details of the fragment's  decorations are relevant if the search is finalized to find traces of an artist's style in objects belonging to different collections. The visual search engine should be therefore designed with sufficient flexibility, giving the possibility to the users to tune and drive the search during their reasoning processes.

The reminder of the paper is organized as follows. In Section \ref{sec:casestudy} we describe the case study in which the similarity questions of this paper take place. In Section \ref{sec:methods} we present the modeling that we adopted to address similarity evaluation, according to the requirements of the user group of the GRAVITATE project and to the characteristics of the fragments. 

In Section \ref{sec:comparison}, we describe how the search engine was conceived as a combination of several the library of descriptors and how they act as independent filters on fragments. Finally, in Section \ref{sec:conclusion}, we draw some conclusions and sketch a plan for future development.

%The translation in algorithmic terms of the similarity criteria that are of interest in the CH domain, 

%Within the dataset, we distinguish between elements that are most \emph{surfacic}-like from those that have a shape per-se; we name them \emph{sherd} or \emph{non-sherd} like.

%The search engine functioning ??? descriptor names, sliders, and results extraction  ??? have been re-thought to better capture the shape similarity search goals that seem to be interesting to the users.
%

\section{Similarity in the archaeological context}
\label{sec:casestudy}
Archeology is a challenging context for similarity assessment for various reasons. First of all, similarity is the guiding reasoning process for the analysis of artifacts and it involved all aspects which may contribute to the shape: geometry, color, textures, chemical properties of color pigments, and many more. Beside data related to the tangible essence of assets, artifacts come with all the contextual information, primarily in textual form, related to the archaeological description and documentation of the findings themselves. 

What complicates the computational model of similarity is the fact that, typically, assets are fractured, eroded, and their color faded. Moreover, pieces of fractured objects may be mixed-up. At the same time, there are a lot of visual cues that may suggest that fragments belong to the same object, and these cues are used jointly or in sequence to answer typical questions related either to re-assemblability (\textit{may these fragment match together?}) or to re-association (\textit{could this fragment belong to that collection?}) or simply to the stylistic analysis of the fragment itself (\textit{do these fragments share the same decoration}?).

In the initial GRAVITATE project phase, we worked with the archaeologists to understand what were the basic similarity criteria used in traditional studies, within the scope of the project case studies. The first part of this section is dedicated to the criteria we identified in the user requests, while the second part describes which datasets are used for our experimentations. 

\subsection{Similarity characterization}
%The similarity criteria act as filters, whose behavior takes into account the type of the query (sherd or non-sherd) and the fragment properties. We have identified the following criteria as guidelines for the development of tools for the similarity reasoning.
The properties listed below synthesize the criteria around which archaeologist build their reasoning about the possibility that fragments belong to the same object, or belong to the same collection. Remember that, in general, re-assemblability can be assessed in very few cases as most of the pieces do not fit together sharply, and several pieces may be missing.

\begin{enumerate}
\item \textit{Overall fragment size}: the size of a fragment, in terms of the overall space occupied by the fragment. Alone, such a criterion is generally too rough to assess similarity, but it becomes useful for the re-assembly of an object, when the elements in place suggest a specific size for the missing fragment(s). 

\item \textit{Thickness}: to be assembled together, sherds should have the same thickness, measured usually along the fractured part of the fragment. 

\item \textit{Material texture}: the interior part of terracotta statues provide interesting insights as similar material roughness may identify a similar manufacturing workshop, for instance a specific mold or chisel, used to produce the original artifact.

\item \textit{Shape continuity}: if two fragments fit close together on a same object, it is likely that they exhibit a similar, overall curvature of the outer skin surface. This criterion is suitable for sherd-like fragments.

\item \textit{Color}: painting over the fragments provide a rich set of visual cues for assessing similarity, and color distribution is an important one. This criterion acts as a rough filter to group fragments of a specific color or material (for instance a stone) independently of the details of the decorations.

\item \textit{Decorations}: some fragments exhibit parts of the same decorations, usually as local relief or colored/painted patterns, providing again an important insight for similarity assessment. 

\item \textit{Overall shape}: frequently, statues are found in pieces which represent a specific anatomical part of the statue; for these fragments, it is useful to consider \textit{global} shape similarity. This criterion is the closest to the classical global similarity criterion but it suitable only for fragments that are a well precise shape (eg, statue's heads).
\end{enumerate}

The criteria listed above represent the basic similarity components, or comparison axis, that are used by experts as single searches or as concatenated filters in their manual analysis tasks. As it will be shown later on, these criteria are used with different weights on different parts of the fragments, and mixed in various manners. 

%different parts of the fragments Finally, as suggested by the experts, since most fragments have an internal and an external facet (e.g.: a fragment of a pot) some of these criteria might need to be considered only on a particular area of the models.

\subsection{Dataset characterization}
The primary GRAVITATE collection is composed by fragments of terracotta figurative statues discovered in Salamis, on the island of Cyprus, dating back to the seventh - early sixth century BC \cite{vassos1991}. Most of these statues are fractured, shards are faded and eroded: in the project, we worked on 241 digital models of Salamis  statues fragments. 

Another interesting case study is represented by the Naukratis collection at the British Museum. The peculiarity of this dataset is that curators provided over the years a series of rich comments, identifying fragments that belong to the same object and documenting the reasons for these conclusions. Those comments also reveal contradictions between curators and inconsistencies in the data. Based on the curator's conclusions, fragments that are likely to belong to the same object are grouped together, but they do necessarily join physically. The collection in our study is made of 72 digital scan of the fragments.

\begin{figure*}[ht]
\includegraphics[width=12cm]{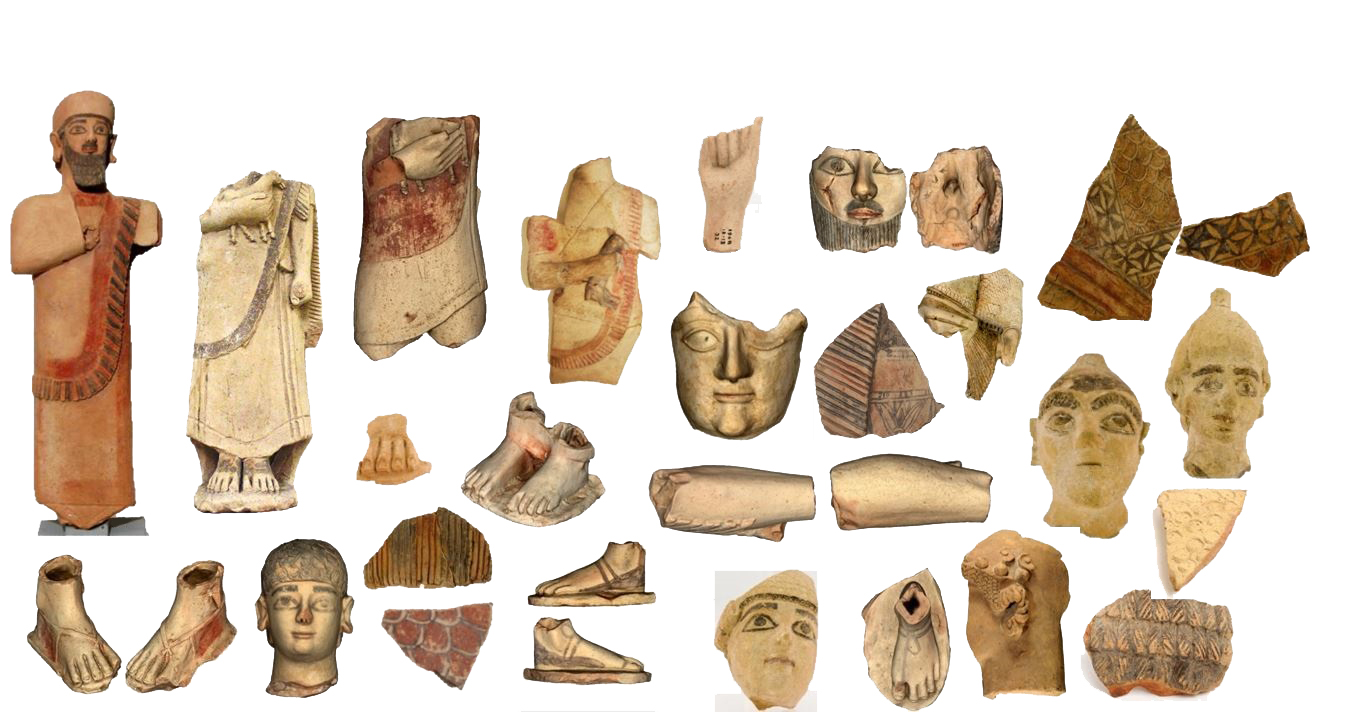}
\caption{Examples of fragments of Salamis terracotta figurines.}
\label{fig:datasetS}
\end{figure*}

\begin{figure}[ht]
\includegraphics[width=\textwidth]{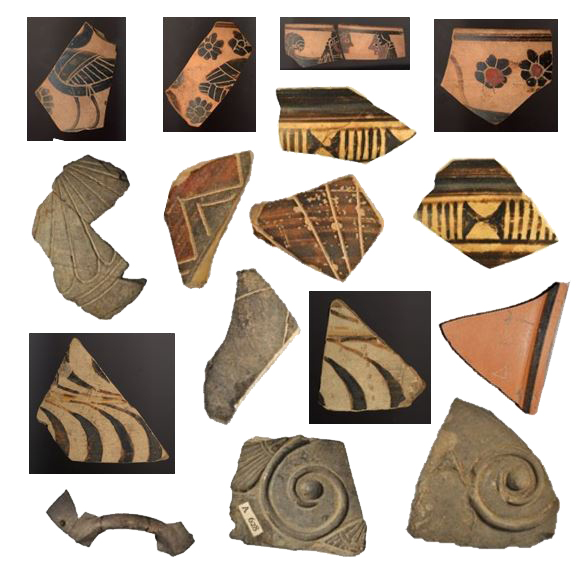}
\caption{Examples of fragments from the Naukratis collection.}
\label{fig:datasetN}
\end{figure}

Figures \ref{fig:datasetS} and \ref{fig:datasetN} show some examples of the fragments of our collection. 
\section{Conceptual modelling of the similarity engine in GRAVITATE}
\label{sec:methods}
In the field of 3D model search and retrieval, a plethora of techniques have been developed to search for 3D objects according to similarity \cite{TangelderV08,bustos05,Biasotti2016:CGF} or for part-based search and retrieval in 3D object databases \cite{Liu13,Savelonas2015} or specific contexts, like product design \cite{SurveyProduct,IYER2005}.
%Besides a single similarity score, .

When reasoning about similarity in a collection of fragmented artifacts, however, there are several  aspects of interest and novelty with respect to the traditional 3D object retrieval operations. 
For instance, the nature of CH data calls for methods dealing with multimodal information in combination (e.g., geometry and texture), which is necessary to effectively group artifacts or their parts into meaningful clusters. This is partially in contrast with the current scenario: many state-of-the-art methodologies for comparing, retrieving, or classifying objects
in repositories are based on a single analysis of the geometric 3D shape \cite{SfPr*14}, possibly building on specific geometric requirements such as the presence of axis of symmetry \cite{KoPa*10} or appendages \cite{KoCh11}. 
Even if several recent approaches for similarity assessment aim at identifying (dense or sparse) correspondences among the model elements (e.g., \cite{OvBe*12,Kim:2012,meshHOG,Smeets2013}) or combining texture and geometry information (e.g., \cite{JOCCH,Biasotti2016,garro16}, they actually pursue a global shape matching approach rather than evaluating similarities based on the comparison of specific features, as necessary when reasoning with archaeological fragments.

Another very relevant aspect relates to the identification of criteria to assess shape compatibility among fragments as a support to the formulation of re-assembly hypothesis. For instance, the pipeline for fragment re-assembly and completion  in \cite{Papaioannou:2017} proposes a number of strategies that range from the fragment faceting to the identification of global feature lines; however, it seems well suited for specific classes of objects like architectural and vessels artifacts. Therefore, most of the symmetry assumptions are not suitable for statue fragments as well as the adoption of methods tailored for articulated objects \cite{Lian11,Lian:2013}.

Based on the intuitive notion of similarity expressed by users (see Section \ref{sec:casestudy}) we describe in the following the shape descriptors that were adopted to compute similarity according to each criterion listed, and discuss the modelling of the fragments that we considered to apply the descriptors properly.

% Besides re-assembly, it is also possible to re-associate two objects that are candidate to have the same origin, for instance, because potentially made with the same mold or created by the same artist.
% In this scenario, the overall shape is probably the less relevant cue to search within collections, while skin decorations contains lots of interesting insights; the partial similarity problem becomes therefore crucial to be solved not only in terms of part-in-whole matching.

\subsection{Matching similarity criteria to signatures}
\label{sec:matching}
% The similarity criteria act as filters, whose behavior takes into account the type of the query (sherd or non-sherd) and the fragment properties. We have identified the following criteria as guidelines for the development of tools for the similarity reasoning.

Starting from the criteria identified to assess similarity for archaeological fragments, we have translated them into computational tools, either considering state of the art descriptors or implementing new methods. Since this is an ongoing work, in this paper we present only the results for all similarity criteria but skin-continuity and pattern recognition. In the following, we will be using the term \textit{compatibility} instead of \textit{similarity}.

\begin{enumerate}

\item Compatibility in terms of \textit{overall size}. In the tests, we have translate this criteria into the comparison of the diagonal of the minimal bounding box (MBB) and the minimal bounding box aspect ratio (MBBAR). Figure \ref{fig:MMBOX} depicts the meaning of the descriptors based on the minimal bounding box. Being scalar values, as distance between the diagonal of the MBB or the MBBAR values, we adopt the  $L^1$ distance.
\begin{figure}
\centering
\includegraphics[width=9cm]{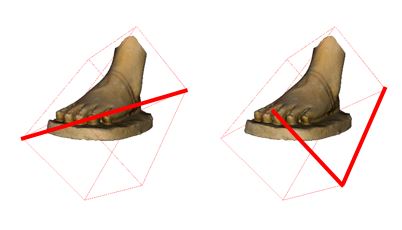}
\caption{Left: the diagonal and Right: the ratio between the minimum and maximum edge (aspect ratio) of the minimal bounding box.}
\label{fig:MMBOX}
\end{figure}

\item Compatibility in terms of \textit{thickness}. The thickness of a 3D model is computed as the shape diameter function (SDF) \cite{Shapira:2008} that has been shown to provide a stable approximation of the diameter of a 3D object with respect to a view cone centered to the surface normal. The shape diameter function is defined for each triangle mesh. Since we are interested in extracting an overall thickness value for the fragment, we consider as thickness descriptor the average of the most frequent value of the SDF. With reference to Figure \ref{fig:thickness}, the thickness value correspond to the abscissa of the maximum value of the histogram. As distance between two thickness values we adopt the usual $L^1$ distance.

\begin{figure}
\centering
\includegraphics[width=9cm]{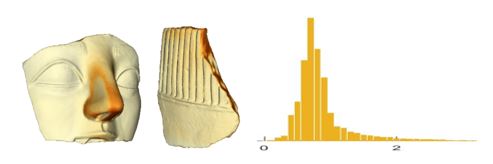}
\caption{Two fragments are colored according to the local SDF value (ranging from red, low values to while yellow, high values. Right: the distribution of the SDF function; the abscissa of the maximum value of the histogram corresponds the thickness value.}
\label{fig:thickness}
\end{figure}

\item Compatibility in terms of \textit{roughness}. To characterize this aspect of fragments, we have identified the mean curvature and the shape index as two possible shape properties. The overall roughness is stored into histograms of the distribution on the surface of the Mean Curvature and Shape Index \cite{Koenderink1992} in a fixed interval. In our settings, we computed the value of the minimum $k_1$ and maximum $k_2$ curvatures on each vertex adopting an implementation of the method \cite{Cohen03}. The mean curvature $K$ and the shape index $SI$ are derived for each vertex, as $K=\frac{k_1+k_2}{2}$ and $SI = \frac{2}{\pi}\arctan\left(\frac{k_1+k_2}{k_1-k_2}\right)$, $k_2\geq k_1$, respectively. An histogram of 200 bins is kept as signature of that surface, see Figure \ref{fig:roughness}. We adopt the Earth Mover's distance \cite{Rubner2000} as the distance between two histograms.

\begin{figure}
\centering
\includegraphics[width=9cm]{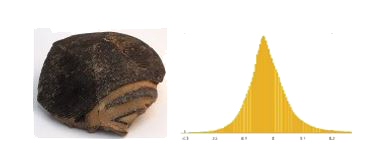}
\caption{The hair is represented on the fragment by a regular relief pattern (left) and the corresponding distribution histogram of the mean curvature is shown on the right.}
\label{fig:roughness}
\end{figure}

%\item Compatibility in terms of skin continuity. \textcolor{red}{non implementata per ora}

\item Compatibility in terms of \textit{color distribution}. We have considered the concatenated histogram of the three color channels (L, a and b) in the CIELab space \cite{Suzuki01,JOCCH}, see Figure \ref{fig:color}. As distance between two color histograms we have considered the Earth Mover's distance \cite{Rubner2000}.  In addition, we have considered also persistence spaces on the color \cite{JOCCH} with similar results, thus we report only the results derived from color histograms.

\begin{figure}
\centering
\includegraphics[width=9cm]{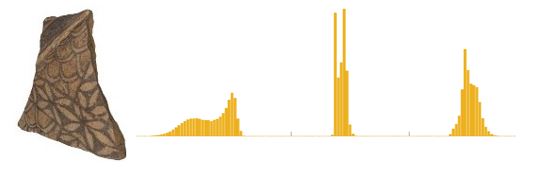}
\caption{A fragment and the corresponding, concatenated color histogram with respect to the L, a and b channels.}
\label{fig:color}
\end{figure}

%\item Compatibility in terms of 2D and 3D patterns \textcolor{red}{Ci lanciamo con qualche matrice?}.

\item Global \textit{overall shape} similarity:  Due to the large variability of shapes within the collection (heads, feet, legs, busts, broken arms and hands), the different scale of the artifacts (statues are often at different scales), the presence of cracks and missing parts, the definition of a geometric signature is an hard challenge. The quite limited number of elements per class also prevents the use of learning techniques. At the end, we opted for a combined shape description that mixes rough filters with scale invariant and non-rigid descriptions. Namely, we adopted a combination of multiple global shape descriptors, like compactness and the minimal bounding box aspect ratio \cite{coarsefilters}, the spherical harmonics indices \cite{SH}, a non-normalized variation of the shape distributions \cite{Osada:2002} and the persistences spaces computed according to the average geodesic distance and the distance from the main axis of an object \cite{BiCe*11}. Figure \ref{fig:shape} represents two histograms corresponding to the non-normalized distribution on the shape of the $D_2$ chord descriptor and the persistence diagram with respect to the average of the shape geodesic distances \cite{hilaga}. All the distances adopted in this setting are metrics in the descriptors space. 

\begin{figure}
\centering
\includegraphics[width=9cm]{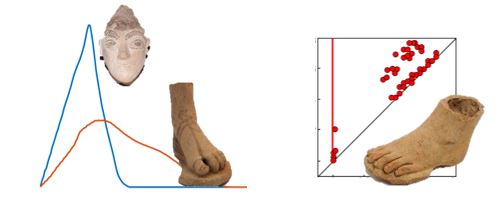}
\caption{Left: The non-normalized $D_2$ signatures of two fragments; Right: the persistence diagram with respect the average, geodesic distance of a fragment of feet.}
\label{fig:shape}
\end{figure}

\end{enumerate}

\subsection{Modelling fragments}
The Salamis 3D models are acquired by laser scans, while the Naukratis models are obtained with photogrammetric scans. All these 3D models are part of the GRAVITATE use cases \cite{GRAVITATE_prog}. Each 3D model is represented as a 3D mesh, equipped with colorimetric information on the vertices.

Looking at the models, it is almost impossible to use most the common descriptor for objects, mainly for the following two bounded reasons. The first, most of the similar problem faced in the literature have benchmarks on models with less than 20,000 vertices (most original models in the dataset have way more than 1,000,000 vertices): this requires a simplification of most models, which brings to a loss of information (in this case, the model lose the smaller features/details). This brings in the second issue, as diversities between the feature of the fragments are far less accentuated than those in most known benchmarks and lies right in the smaller details. 
Because of this, most of the descriptor methods for objects are discarded a priori.

%\textcolor{red}{We expected that the descriptors for free-form objects be relevant to these artifacts as well, since they are based on elementary geometrical properties that distinguish features in any context: non capisco bene cosa si vuole dire qui @: ci si aspetta che i descrittori di forma siano forti in questo campo, siccome sono basati su feature presenti in ogni contesto (non sono specializzati, in breve). Penso intendesse questo. Nel caso, lascerei così.}. However, the type of similarity needed for the fragments in our tests is more challenging than most of the problems approached in the literature, either in terms of size of the models (most of the benchmarks have models with less than 20,000 vertices) and in the fact that the (partial) similarity must inferred from subparts of fragments or facets.

% As stated before, a crucial aspect when dealing with broken, ceramic artifacts it is the fact that most of fragments are characterized by their decorations and chiselings rather than their overall geometric shape. 

Beside the association of intuitive criteria to signatures and distances, an important issue is that some of the criteria identified implicitly assume a classification of the fragment types into two broad categories: those shaped like a classical sherd and those representing still a volumetric component of the original object. For instance, with reference to Figure \ref{fig:datasetS}, the heads, the figurines and the busts are examples of volumetric fragments  while the decorated fragments (e.g. top-right and bottom-right) or the fragments in Figure \ref{fig:datasetN} are examples of sherd-like fragments. In the discussions with the experts, global shape similarity was actually indicated as a criteria for evaluating similarities among volumetric fragments, while for sherd-like fragments global shape similarity is of little utility. Conversely, thickness is primarily associated to the analysis of sherd-like fragments, while the concept of thickness itself is more less useful to compare more volumetric fragments.

For this reason, we performed a first distinction between fragments shaped like \textit{sherds} and those having an almost volumetric shape (named non-sherd-like). This distinction is fundamental for the functioning of the search engine.

%Descriptors are now tailored to the type of fragments: sherd-like vs non-sherd-like  (classification done with the support of Cyprus Institute)
Another important issue is that the intuitive criteria are applied, implicitly, to portions of the fragments only: for instance, the shape continuity is actually thought for the outer surface of the fragment, the one which correspond to the exterior surface of the statue or vase. In GRAVITATE, we have therefore modelled the 3D meshes representing the fragments as a set of \textit{facets}, corresponding to the skin, fracture or interior part of the original object. In \cite{Faceting}, the challenging task of decomposing automatically the fragment surface into facets is discussed. 

The initial results presented in this paper do not use localization of descriptors into facets, which is currently being developed and integrated in the GRAVITATE search engine. 

\section{Fragment comparison}
\label{sec:comparison}
%% EDIT intro search Engine: 
The proposed fragment comparison framework acts as a query-by-example search engine. Starting from a model $A$ (query) of a given dataset $R$, a list $L$ of models in $R$ (could be empty) that are considered similar to the query model $A$ is retrieved. The search is based on the activation of a number of properties ($P_1$, $P_2$,...) that the user can select before running the search, via check-boxes. Each property corresponds to one of the search criteria discussed in Section \ref{sec:matching}. The type of query object (sherd/non-sherd) determines what check-boxes are enabled, that is, what properties are considered valid search criteria based on the nature of $A$ (e.g.: if $A$ is sherd-like, the filter related to properties that works on non-sherd-like object are not activated).

Each property $P_i$ is interpreted as a \emph{filter} ($F_i$), which removes from the results list $L$ the models that are dissimilar from $A$ based on $P_i$. If the filter results too restrictive based on the user judgment, it is possible to relax the filter severity, which incrementally add more models to $L$ gradually less similar to $A$ (with respect to the filters selected).

From a more technical point of view, the filters are distance matrices among models, one matrix for each property. 
When the user selects one or more criteria to compare a query model against the dataset, each criterion acts as a filter and the combination of more filters is done on the basis of the logic $and$ operator. In practice, the model $A$ is similar to the model $B$ with respect to the two properties $P_1$ and $P_2$ if the two models $A$ and $B$ are similar with respect to both $P_1$ and $P_2$. Note that the combined distance is a metric if all the distances for the single properties are metrics.

The query results that fulfill all the criteria are ranked according to a combined measure defined as the product of all the distances that are smaller than a threshold: this threshold acts as a tuning parameter for the granularity of the search results, using the criteria described below.

A different threshold is set for each shape property, as follows. Given a distance matrix $D$, the $k-th$ row $Q_k$ stores the distance of the $k-th$ model with respect to all the other elements in the dataset. Given a threshold $t$, the set $V_k$ of models such that $Q_k(j)\leq t$ are considered valid query results for the model $k$ with respect to the property stored in $D$.

Since a similarity between models is not binary (similar or not similar) but it is defined by a similarity measure, it is impossible to set thresholds a priori, thus they are defined as follow. 
For each matrix $D$, the threshold $t$ is automatically predetermined as:
$t =average_k\{t_k\}$,  where $t_k=  argmin\{ \#\{V_k\} $ 5\%  of the elements of the dataset\}. This approach allow the user to find the most similar model to the query model with respect also to the variety of the dataset.

By tuning the thresholds values, it is also possible to enlarge the search result set, offering more flexibility and interaction to the user. A value $dt$ is determined as the average of the $dt_k$ values that increase the cardinality of $V_k$ of approximately 3\%. The value $dt$ is selected as many times the user decides to relax the threshold $t$. The thresholds values $t$ and $dt$ are automatically determined for each property. 

\begin{figure}
\centering
\begin{tabular}{|c|}
\hline
\emph{Color}\\
\includegraphics[width=10cm]{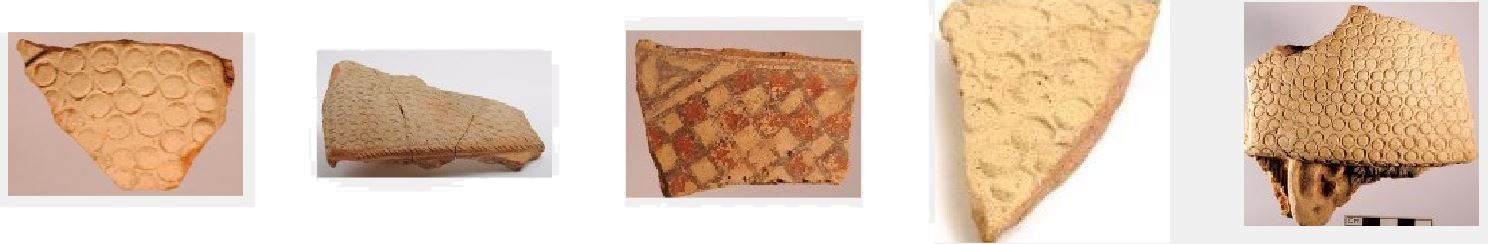}\\
\hline
\emph{Color}\\
\includegraphics[width=10cm]{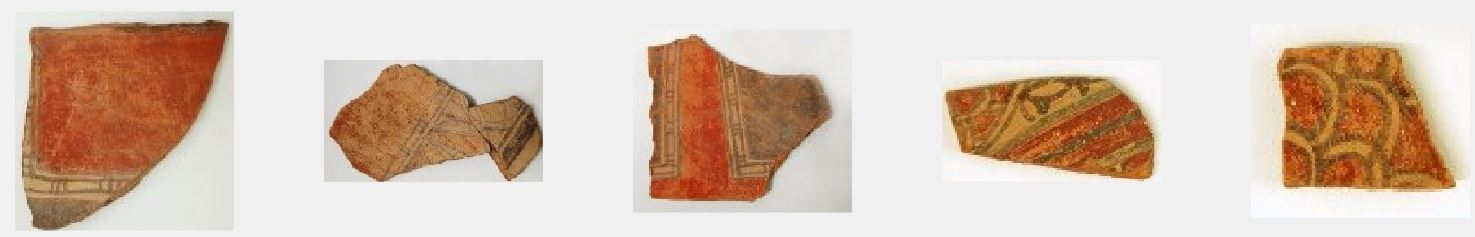}\\
\hline
\emph{Color}\\
\includegraphics[width=10cm]{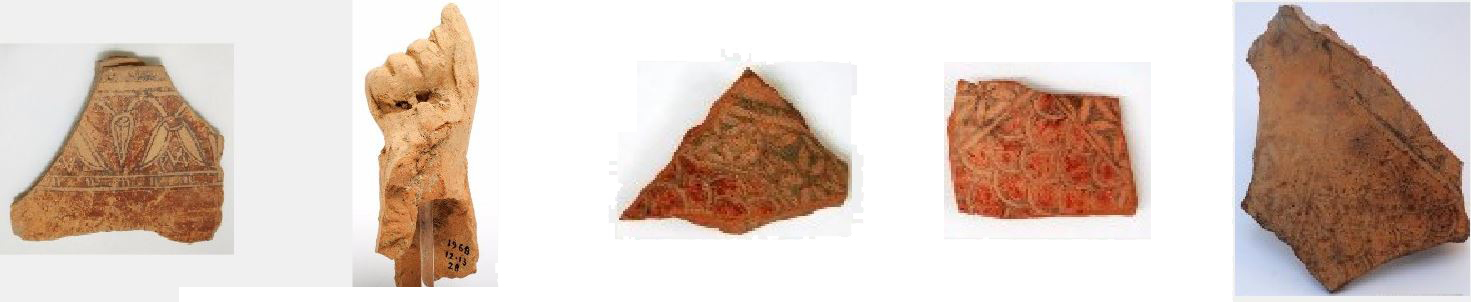}\\
\hline
\emph{Color+Thickness}\\
\includegraphics[width=10cm]{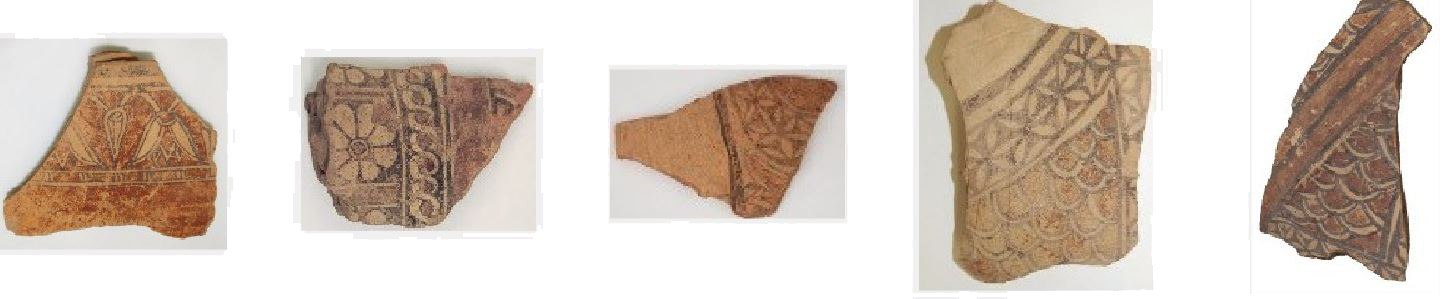}\\
\hline 
\end{tabular}
\caption{The query is the leftmost model of each row. First three rows: examples of results by color histograms. Last row: the same query of the 3rd row, refined also by thickness.}
\label{fig:colorR}
\end{figure}

\begin{figure}
\centering
\begin{tabular}{|c|}
\hline
\emph{Color+Thickness}\\
\includegraphics[width=10cm]{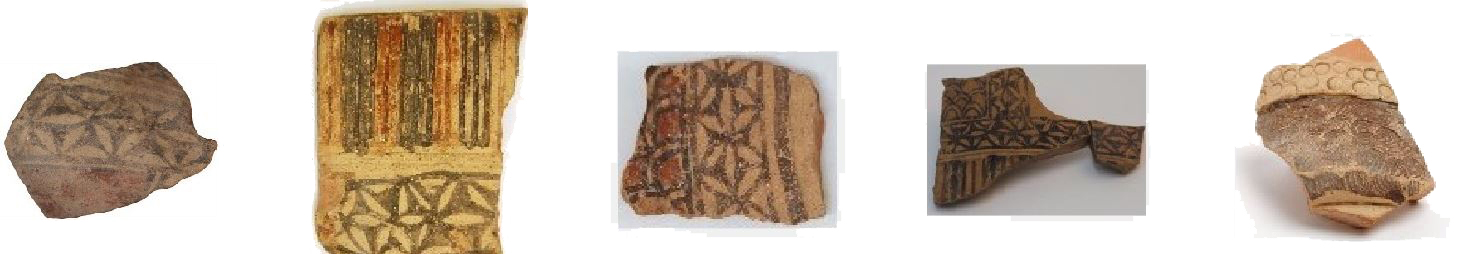}\\
\hline
\emph{Color + Thickness + Roughness}\\
\includegraphics[width=4.5cm]{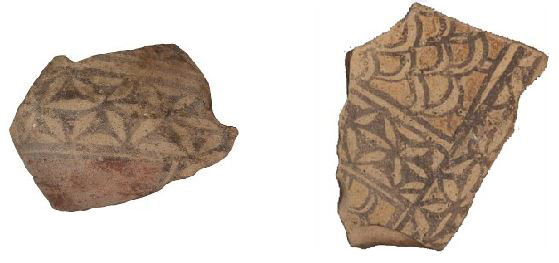}\\
\hline
\emph{Color + Thickness + Roughness, Relaxed}\\
\includegraphics[width=10cm]{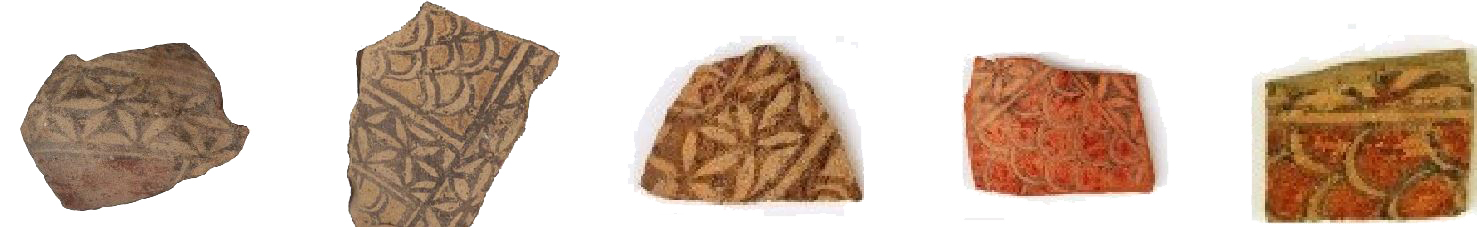}\\
\hline
\end{tabular}
\caption{An example of the application of multiple filtering and relaxation.}
\label{fig:colorTR}
\end{figure}

In the following we discuss our experiments on similarity reasoning. The validation of the query results was done visually, discussing the outcome with respect to positive and negative examples proposed by the CH experts. 
Figure \ref{fig:colorR} shows some examples of query results when the concatenated color histograms are adopted as the descriptor. In this picture, the query model is the leftmost model of each row. In the first row, the query object is characterized by circles; the square-like fragment in the middle is a false positive. The query object in the second row is characterized by a large red band with lateral decorations; therefore, the scales-like decoration in the rightmost position is considered by the expert as a false positive example. In both cases, we expect that adopting a pattern description would solve these false positive examples. As an example of how the combination of multiple criteria can overcome some limitations, in the forth row we show how the adoption of a double filter for color and thickness, it is possible to select only decorated fragments that are potentially compatible.
Similarly, Figure \ref{fig:colorTR} shows how multiple filters act. For instance, when applied to a quite smooth fragment like the leftmost in the first row of Figure \ref{fig:colorTR}, the roughness filter discards fragments that are a jagged decoration on the external skin, even after the relaxation of the similarity criteria.  When, %our are promising, as using a single filter groups models that are considered similar from CH experts. 
the use of multiple filters seems to be a bit restrictive (Figure \ref{fig:colorTR}(Middle row)), the relaxation option allows to add more similarity options (based on the selected  criteria).

\begin{figure}
\centering
\begin{tabular}{|c|}
\hline
\emph{Color+Relaxed}\\
\includegraphics[width=10cm]{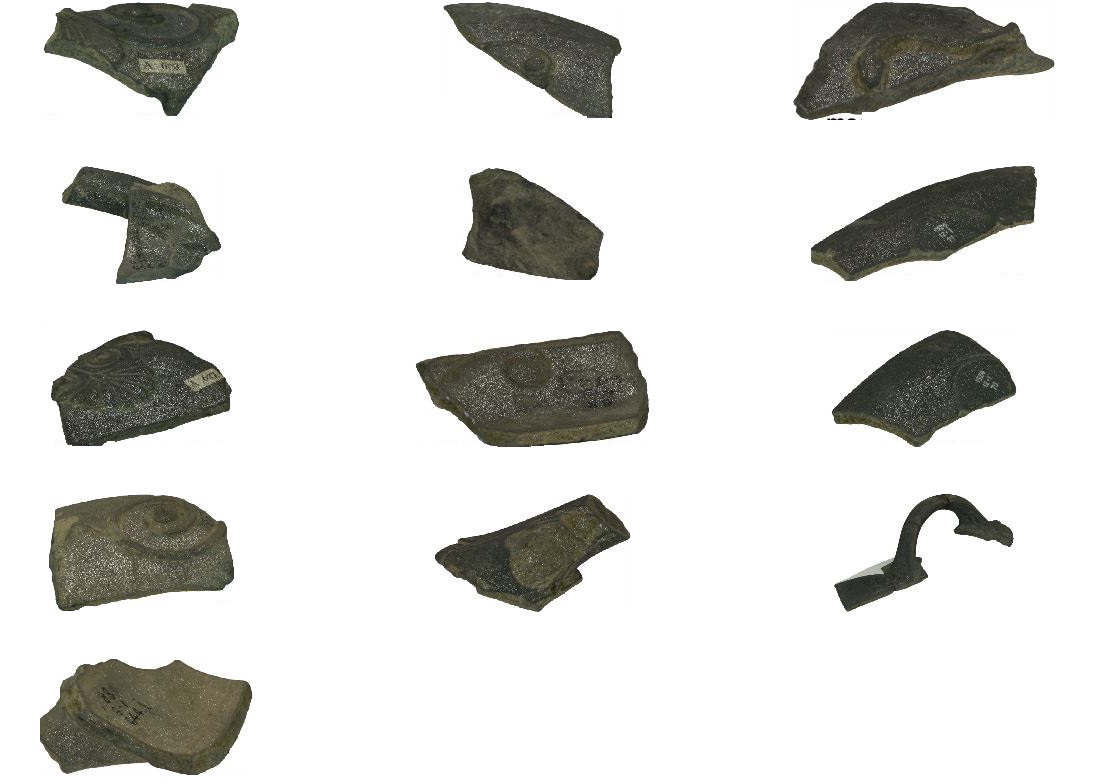}\\
\hline
\end{tabular}
\caption{Example of query results using the top, leftmost model with respect to the color filter.}
\label{fig:colorN}
\end{figure}

\begin{figure}
\centering
\begin{tabular}{|c|}
\hline
\emph{Color}\\
\includegraphics[width=10cm]{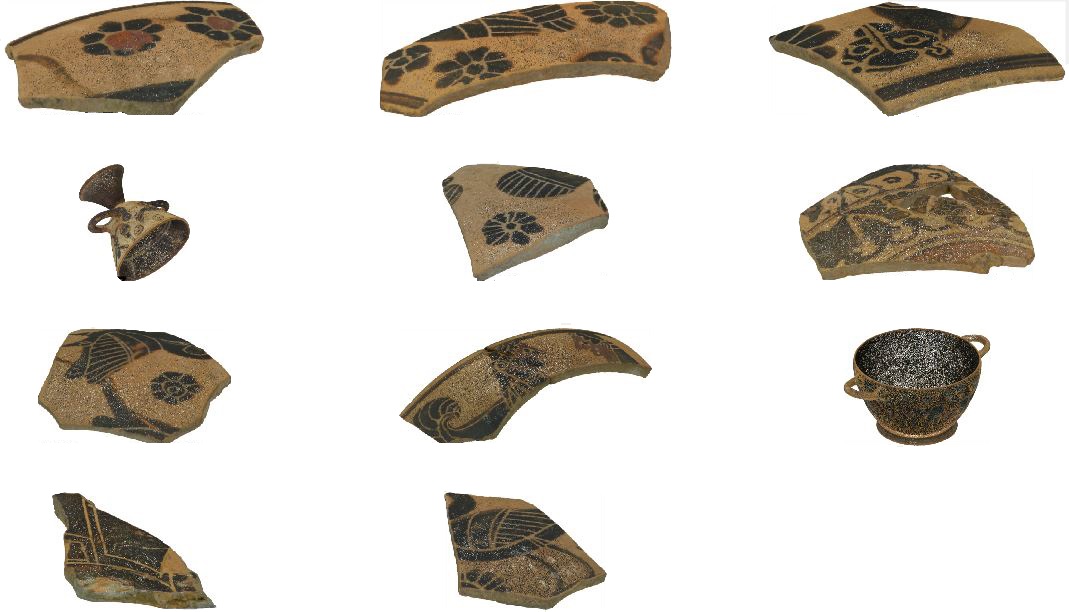}\\
\hline
\emph{Color + Thickness}\\
\includegraphics[width=10cm]{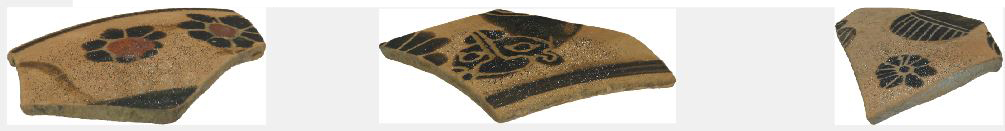}\\
\hline
\end{tabular}
\caption{Example of query results using the top, leftmost model with respect to the color filter (Top row) and color and thickness filter (Bottom row).}
\label{fig:colorNN}
\end{figure}

Other examples of color matchings are shown in the Figures \ref{fig:colorN} and \ref{fig:colorNN}. In both cases, the fragments come from the Naukratis collections and are grouped by the curators according to their provenance. In the example in Figure \ref{fig:colorN}, using only color and relaxation on the automatic color threshold, we are able to select all the fragments in the dataset that belong to the same group. One of the peculiar characteristics of this group is that all its elements are made by a stone. In the example in Figure \ref{fig:colorNN}, the query results obtained with the color filter contain two false positives: the complete vase in the leftmost position in the second row and the fragment in the rightmost position in the second row. Anyway. the CH experts judged that these false positive elements are stylistically compatible with the elements of the group. The bottom row highlights the fragments that are also compatible by thickness.

\begin{figure}
\centering
\begin{tabular}{|c|}
\hline
\emph{Shape}\\
\includegraphics[width=10cm]{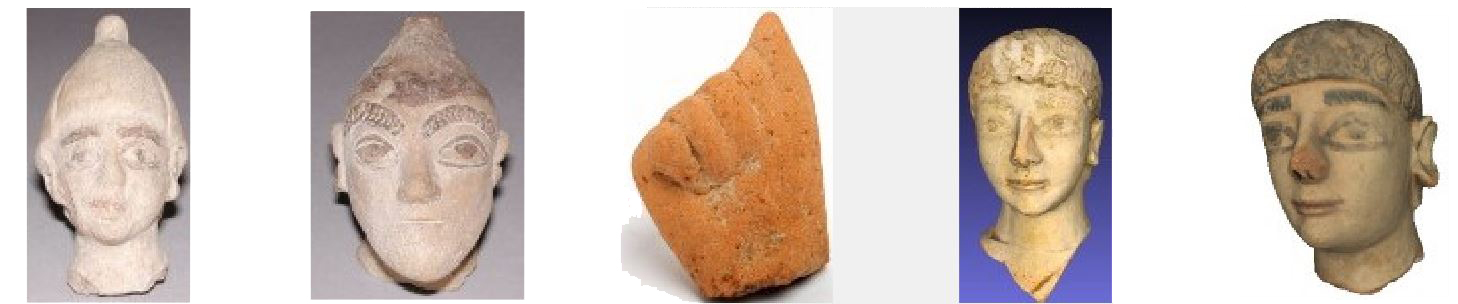}\\
\hline
\emph{Shape}\\
\includegraphics[width=10cm]{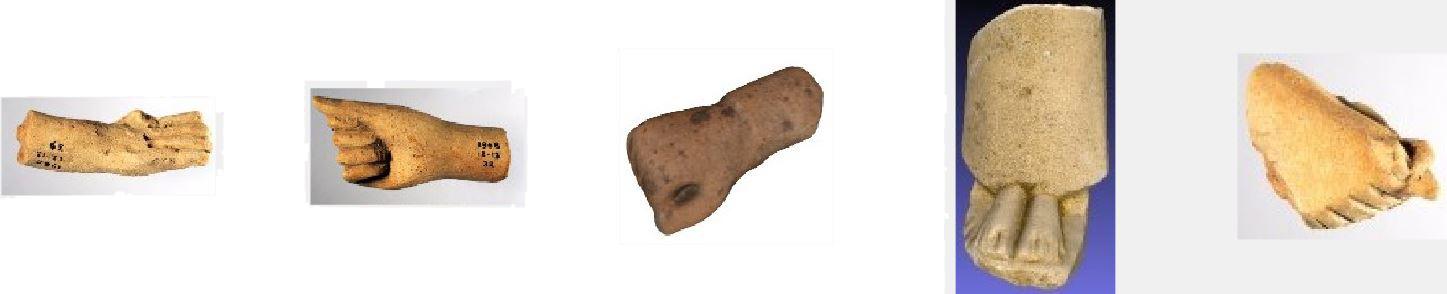}\\
\hline
\emph{Shape + Envelope}\\
\includegraphics[width=4.5cm]{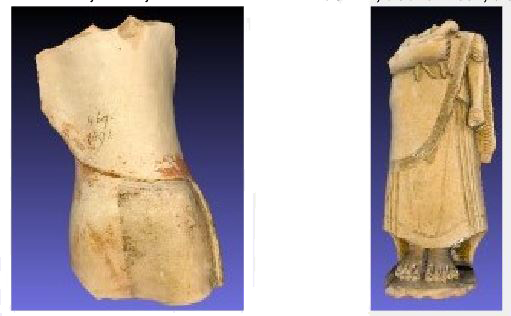}\\
\hline
\end{tabular}
\caption{Examples of similarity searches using global properties of the query models.}
\label{fig:shapeS}
\end{figure}

Addressing the overall shape matching problem is really challenging. When dealing with broken fragments of statues there is a large intra-class variability, often emphasized from different points of fracture. For instance, the class of \emph{hands} contains complete and broken hand-models but also parts of forearms and fists. Similarly, belonging to different statues, heads can have different scale and decorative elements, like helmets, that influences the overall geometry of the fragment. Finally, the descriptors in use for describing the overall shape of a 3D model provide a quite rough filter, discarding the details and any semantic meaning associated to the shape. Figure \ref{fig:shapeS} shows some of the query results on the dataset of non-sherd fragments using the combination of shape descriptors as filter. Both examples in the first row and in the second row obtain a false positive result (the middle element in the first row and the fourth elements in the second row). From our perspective, these false positive outcome derive form the use of global shape descriptions that mainly consider the overall shape embedding rather its details, see for instance the head with the helmet and the fist examples. Finally, in the third row we show how the combination of shape and envelope (described in terms of the diagonal of the minimal bounding box) restricts the query results to statue fragments that have approximately the same size.

These results of these preliminary experiments show that computational methods have the potential of supporting the CH experts with quantitative estimations of their findings but we are still far from having a satisfactory solution to the similarity assessment problem in the CH domain.

% %\subsection{Analysis of the similarity}
% \textcolor{red}{Metterei tante figure di esempi visuali. Come ground truth possiamo mettere l'analisi di quanto i descrittori scoprono gli sherd non sherd ma non mi sembra molto utile, tanto pi che poi l'abbiamo inglobata come conoscenza pregressa.}

% \textcolor{red}{Questo vuol dire che io metterei almeno una figura per criterio}
% %
% %\subsection{Discussions}
% \textcolor{red}{Qui, riprenderei i punti dei criteri e dei tool ssviluppati per discutere punto per punto cosa funziona e cosa c'è da migliorare... dubbio se questa e la precendente non siano la stessa sezione, forse si, altrimenti si rischia di avere la carrellata di figure e poi i commenti}.
% %
\section{Conclusions}
\label{sec:conclusion}

Working with real use cases poses a lot of problems, which span from the need of keeping results consistent across different resolutions to the management of damaged fragments. A peculiar aspect of the archaeological fragments is that they do not have a unique interpretation and multiple similarity criteria can be applied depending on the archaeological interests. 
The classical partial matching problem interpreted as a part-in-whole problem is not enough to address similarity among fragments because the original model is generally missing and the archaeologists only own small fragments that cannot be completely reassembled. 
In this sense, it is necessary to develop forms of multi-signatures that specifically target the different fragment facets; for instance, the external facet for decorations and internal one for recognition the print of the technical turning.  

Many problems are still open and need further efforts to be solved.
For instance, skin continuity, in terms of compatible overall skin curvature, is crucial for re-assembling artifacts. Currently, it is addressed for models with symmetries, such as potteries, for instance considering the the partial axial symmetry of the surface \cite{sipiran_ICCV}. 
Moreover, the CH artifacts present quite complex features such as decorations, style elements and patterns (either colorimetric and geometric) that require the development of specific descriptors. 

We are currently working on localizing the descriptors to facets only and on the integration of further descriptors into the GRAVITATE search engine.

As mentioned in Section \ref{sec:matching}, in the CH domain  there is a lack of substantial training data that in our opinion partially prevent the adoption of learning techniques. Indeed, learning techniques are based on supervised information, requiring extensive training data in addition to practical configuration expertise and computational resources (Lecun et al. 2015), (Boscaini et al. 2015). 
On the contrary, in our use cases only a few shapes of a certain class or type exist, and they do not correspond to existing, complete shapes. Using methods like those addressed in this paper, a domain expert might be better able to determine the type of properties needed (occurrence of feature points, curvature statistics, repeated patterns, etc.) and which may be interpreted to some extent.

\section*{Acknowledgments}
The work is developed within the research program of the ``H2020'' European project ``GRAVITATE'', contract n. 665155, (2015-2018).

\bibliographystyle{abbrv}
\bibliography{bibliography}
%-------------------------------------------------------------------------
\newpage

\end{document}